\documentclass{article}
\usepackage{spconf,amsmath,graphicx, multirow, url, float, booktabs, amssymb}

\title{Interpreting glottal flow dynamics for detecting COVID-19 from voice}
\name{Soham Deshmukh, Mahmoud Al Ismail, Rita Singh}
\address{Carnegie Mellon University, Pittsburgh, PA, USA \\
         \texttt{\{sdeshmuk,mahmoudi,rsingh\}@andrew.cmu.edu}}
\begin{document}
%\ninept
%
\maketitle
\begin{abstract}
In the pathogenesis of COVID-19, impairment of respiratory functions is often one of the key symptoms. Studies show that in these cases, voice production is also adversely affected -- vocal fold oscillations are asynchronous, asymmetrical and more restricted during phonation. This paper proposes a method that analyzes the differential dynamics of the glottal flow waveform (GFW) during voice production to identify features in them that are most significant for the detection of COVID-19 from voice. Since it is hard to measure this directly in COVID-19 patients, we infer it from recorded speech signals and compare it to the GFW computed from physical model of phonation. For normal voices, the difference between the two should be minimal, since physical models are constructed to explain phonation under assumptions of normalcy. Greater differences implicate anomalies in the bio-physical factors that contribute to the correctness of the physical model, revealing their significance indirectly. Our proposed method uses a CNN-based 2-step attention model that locates anomalies in time-feature space in the difference of the two GFWs, allowing us to infer their potential as discriminative features for classification.  The viability of this method is demonstrated using a clinically curated dataset of COVID-19 positive and negative subjects.
% In the pathogenesis of COVID-19, the impairment of respiratory functions is often one of the key symptoms. Studies show that in these cases, voice production also adversely affected -- vocal fold oscillations are asynchronous, asymmetrical and more restricted during phonation. To find the signatures that are significant for classification, however, better methods of comparative analysis are required. This paper proposes one such method that analyzes the dynamics of the glottal airflow during voice production to detect anomalies. Since it is hard to measure this directly in COVID-19 patients, we infer the glottal flow waveform (GFW) from recorded speech signals and compare it to the GFW computed from a physical model of phonation. For normal voices, the difference between the two should be minimal, since physical models are constructed to explain phonation under assumptions of normalcy. Higher differences implicate anomalies in the bio-physical factors that contribute to the correctness of the physical model, revealing their significance indirectly. Our proposed method uses a CNN-based 2-step attention model that locates anomalies in both time and feature space within the differences of these GFWs, allowing us to infer the most discriminative aspects within them for use with COVID-19 detectors. The viability of this method is demonstrated using a clinically curated dataset of COVID-19 positive and negative subjects.
\end{abstract}
\section{Introduction}
\label{sec:intro}

The production of voiced sounds is a two-step process involving the self-sustained vibrations of the vocal folds, and the propagation of the resultant pressure wave within the vocal tract \cite{titze2008nonlinear}. In this paper, we focus on the first step: the motion of the vocal folds. Recent studies have shown that this motion is adversely affected in symptomatic COVID-19 patients \cite{ismail2020detection}. While these studies are able to identify broad-level anomalies (such as asynchrony, asymmetry of motion, reduced range of oscillation etc.) by visual comparisons between oscillation patters of healthy and \textit{symptomatic} COVID-19 positive people, more systematic methods of comparison are needed to  identify discriminative signatures that may be used as features in classifiers that aim to detect COVID-19 from voice.

This problem may be approached by first noting that that the motion of the vocal folds is highly correlated with the size of the glottal opening. The glottis opens and closes as the vocal folds vibrate. In this self-sustained cycle of the opening and closing of the glottis, air pressure periodically builds up and is released, giving rise to a fluctuating volume of airflow across the glottis that comprises the glottal flow waveform (GFW). Since the glottal opening and closing is caused by the motion of the vocal folds, the spatial displacements of the vocal folds are normally highly correlated to the GFW. Simple physical models of phonation express these displacements as a function of the bio-physical properties of the vocal folds, such as their thickness, elasticity etc. and of the aerodynamic forces across the glottis e.g. \cite{lucero2013modeling, ishizaka1972synthesis, yang2011computation, alipour2000finite, titze1988physics}. Given a set of parameters, these models yield the oscillation of the vocal folds. 

Ideally, if we were to compute the difference in the GFW of the same speaker when healthy and when symptomatic with COVID-19, we would learn a lot from the differences we observe. However, such pairs of recordings are not easily available. In lieu of this, if only a recording of a COVID-19 positive individual is available, we can compute the GFW (denoted $GFW_{filter}$) from it by inverse filtering the speech signal. In this work we assume that we do \textit{not} have a recording of the same speaker when healthy. We now \textit{deliberately} assume that a physical model that explains the vocal fold motions of a healthy individual will match the those of the COVID-19 positive speaker obtained from inverse filtering. Clearly, this assumption is not true, and will be invalidated to different degrees, depending on how much it deviates from reality. Due to the correlation of the GFW to the vocal fold oscillations, these deviations are expected to surface in the form of differences between the GFW calculated from the estimated vocal fold motions from the physical models ($GFW_{model}$) and $GFW_{filter}$. These differences can then be ascribed to  the influence of COVID-19, which changes the underlying bio-physical factors that the model attempts to emulate, and  renders the assumptions invalid in proportion to the severity of the disease in the speaker.

In this paper, we the ADLES \cite{wenbo} algorithm to jointly estimate the vocal fold oscillations and $GFW_{model}$ and compare it to the $GFW_{filter}$ obtained from the speech signal. We propose a method of analysis of the observed differences, to identify the features in it that are relevant to classifiers for COVID-19 detection. The method uses an interpretable CNN model with a 2-step attention pooling architecture to detect the differences (residual) as a function of time. We show that it successfully identifies the traces that are of significance to COVID-19 detection from voice, and leads to better classification performance when tested on a clinically curated dataset of COVID-19 positive and negative individuals. 

%The organization of paper is as follows. Section \ref{sec:related_work} describes related efforts in detection of Covid-19 from voice. Section \ref{sec:proposed_methodology} describes the proposed methodology. Our experiments are described in Section \ref{sec:experiments}, and followed by results and conclusions in Sections \ref{sec:results} and \ref{sec:conclusion} respectively.

\subsection{Related Work} \label{sec:related_work}
The literature on detecting COVID-19 from voice and respiratory sounds is relatively new \cite{tcs_2020overview} and sparse. Literature on using voice, and especially the dynamics of vocal fold motion to detect COVID-19 is even more sparse, with only one paper in existence \cite{ismail2020detection}. To the best of our knowledge, our current paper is the first one that addresses the problem of interpretation of vocal flow dynamics for more accurate identification of the discriminative factors that can be used for classification of COVID-19 from voice.

\section{Proposed Methodology} \label{sec:proposed_methodology}
\subsection{Computing the GFW from a phonation model}
%In the past decades, several physical models that explain the motion of the vocal folds during phonation have been proposed, e.g. \cite{lucero2013modeling, ishizaka1972synthesis, yang2011computation, alipour2000finite, titze1988physics}. 
We use the 1-mass asymmetric body cover model proposed in \cite{1fold_asym}, and using the ADLES algorithm proposed in \cite{wenbo}, compute oscillation of the vocal folds from which an estimate for the glottal flow waveform $GFW_{model}$ can be obtained. To compute the vocal fold oscillations (or the displacements of the vocal folds), the ADLES algorithm solves the \textit{forward} problem of estimating the displacement and velocity of each vocal fold jointly with the \textit{inverse} problem of estimating the parameters of the dynamical system comprising the model. To do this, we must minimize the squared difference between the estimated glottal flow from inverse filtering $GFW_{filter}$, and $GFW_{model}$.
Mathematically:
\begin{align}\label{eq:m1}
    \min &\int_{0}^{T} (u_0(t) - u_0^m(t))^2 dt\\
    \min &\int_{0}^{T} (\tilde{c}d(2x_0 + x_l(t) + x_r(t)) - \frac{A(0)}{\rho c}\mathcal{F}^{-1}(p_m(t)))^2 dt \label{eq:obj_least_squares}\\
    \mathrm{s.t.}\quad & \dot{x}_r + \beta (1 + x_r^2)\dot{x}_r + x_r - \frac{\Delta}{2}x_r = \alpha (\dot{x}_r + \dot{x}_l) \label{eq:x_r}\\ 
    &\dot{x}_l + \beta (1 + x_l^2)\dot{x}_l + x_l + \frac{\Delta}{2}x_l = \alpha (\dot{x}_r + \dot{x}_l) \label{eq:x_l}\\ 
    & x_r(0) = C_r, x_l(0) = C_l, \dot{x}_r(0) = 0, \dot{x}_l(0) = 0
\end{align}
where the Eqns. (3) and (4) jointly correspond to the model. $x_r, x_l$ are the displacements of right and left vocal fold, $\dot{x}_r, \dot{x}_l$ are horizontal vocal fold velocity of right and left vocal fold w.r.t the center of the glottis, $\alpha$, $\beta$ and $\Delta$ are the model's parameters. $\alpha$ is the coupling coefficient between the supra- and sub-glottal pressure, $\beta$ incorporates the mass, spring and damping coefficients of the vocal folds,  and $\Delta$ is an asymmetry coefficient. Here $GFW_{filter} \equiv u_0^m(t)$ is the GFW obtained from inverse filtering, and $GFW_{model} \equiv u_0(t)$ is estimated GFW as a joint solution to the model with ADLES.

Within ADLES, the optimization problem is cast into its dual by using Lagrangian followed by optimization using gradient descent. The process of obtaining the solution for parameters is  explained in detail here \cite{wenbolong}.

\subsection{COVID-19 analysis framework} \label{sec:attention}
We represent each recording as a sequence of sliding windows (frames) $\{X_i\}_{i=1}^T$, where each frame is of duration $\tau$, and the frame shift is $o$. For each of the frames $X_i$, a $GFW_{filter} \equiv u_{i}(t)$ estimate is obtained by inverse filtering and also as a function of vocal fold motion $GFW_{model} \equiv u_{i}^m(t)$. We will drop the subscript $0$ indicating measurement at the glottis from this notation for brevity. Each frame is thus represented by $[u_{i}(t), u_{i}^m(t)]$, and the sequence of audio frames is now  $\{X_i\}_{i = 1}^T = \{[u_{i}(t), u_{i}^m(t)]\}_{i=1}^T \in \mathbb{R}^{2 \times T}$.

We wish to capture the differences between the two GFW estimates in a way that a classifier is best able to use them to discriminate between COVID-19 postitive and negative cases, i.e., we want to estimate the probability $P(y|u_{i}(t), u_{i}^m(t); \textbf{w})$ where $\textbf{w}$ represents the parameters of a parametric model. $y \in \{0,1\}$ is a set of binary labels used with a classifier. The task of estimating $P(y|u_{i}(t), u_{i}^m(t); \textbf{w})$ can be divided into two stages: in the first, a time-distributed pattern detector is trained to find the similarities and differences between $u_{i}(t), u_{i}^m(t)$ at each time step (a frame comprises $N$ time steps). In the second, a pooling mechanism aggregates the outputs of the first stage to yield a single prediction for each frame.  

Let the output of first stage be given by $\textbf{Z}$ and let the learnable feature extractors be represented by $f_1(.)$. The goal of $f_1(.)$ is to learn a mapping such that:
$$f_1:  \{[u_{i}(t), u_{i}^m(t)]\} \mapsto \textbf{Z}$$ 
The aim of the second stage is to reduce $\textbf{Z}$ from stage 1 along both, the time and feature axes to obtain single frame-level prediction $y_i$. This can be thought of as a multiple-instance learning (MIL) setup \cite{MIL} where each window/frame $X_i$ is represented as a bag of $N$ samples: $B_i = (\{x_j\}_{j = 1}^N, y_i)$ where $x_i$ represents individual samples in the frame. This type of MIL formulation has been extensively used in a Weakly Labeled Sound Event Detection (WLSED) \cite{Anurag_WLD} framework. The pooling methods developed for WLSED \cite{gwrp,atrous,soham} are also transferable to our problem of  MIL based COVID-19 detection. We now introduce a pooling function $f_2(.)$ that  learns the mapping:
$f_2: \textbf{Z} \mapsto y$

The 2-step Attention Pooling (2AP) mechanism \cite{soham} yields benchmark performance for WLSED, in the task of determining the contribution of each class at each instance in both time and frequency domains. In the problem setup in \cite{soham}, the output of the first step of attention is $C \times T$ where $C$ is the number of classes and is the result of a weighted combination of the frequency components. In our case, as we have a single class setup (COVID/non-COVID), the output of first step of attention is $1 \times T$, which is a weighted combination of the amplitudes of $GFW_{filter}$ and $GFW_{model}$. This weighted combination of amplitudes ranges from high to low depending on whether the phase difference between the two signals is low or high respectively. Adding an extra feature detector $f_3(.)$ between the two stages allows the model to better capture the entities of interest to us as a function of time. We call this step the Sandwiched 2-step Attention Pooling (S2AP) step. Subsequently, we use S2AP for learning the mapping $f_2(.)$. Formally, first step in S2AP takes $\textbf{Z}$ as input:
\begin{equation}
    \widehat{Z}_{a1} = \frac{e^{\sigma(\textbf{Z}\textbf{w}_{a1}^T + b_{a1})}}{\sum_{i = 1}^F e^{\sigma(\textbf{Z}\textbf{w}_{a1}^T + b_{a1})}}, Z_{p1} = \sum_{i = 0}^F (\textbf{\textbf{Z}}\textbf{w}_{c1}^T + b_{c1}) \cdot \widehat{Z}_{a1}
\end{equation}
\begin{equation}
    Z_x = f_3(Z_{p1})
\end{equation}
\begin{equation}
    \widehat{Z}_{a2} = \frac{e^{\sigma(Z_x \textbf{w}_{a2}^T + b_{a2})}}{\sum_{t = 1}^T e^{\sigma(Z_x \textbf{w}_{a2}^T + b_{a2})}}, Z_{p2} = \sum_{t = 0}^T (Z_x \textbf{w}_{c2}^T + b_{c2}) \cdot \widehat{Z}_{a2}
\end{equation}
where $\sigma(.)$ is Sigmoid function, $\{\textbf{w}_{a1}, \textbf{w}_{c1}, \textbf{w}_{a2}, \textbf{w}_{a3}\}$ are learnable weight parameters and $Z_{p2}$ is $y$ and gives the probability of the voice sample containing traces of COVID-19. Moreover, S2AP retains the benefits of 2AP of making the pooling more interpretable.  Visualization of the normalized attention weights $\widehat{Z}_{a1}, \widehat{Z}_{a2}$ and the output $Z_{p1}, Z_{p2}$ allows us to infer what features and what time steps contribute most significantly to the decisions made by the classifier. 

% \begin{table*}[!t]
% \small
%     \centering
%     % \captionsetup{justification=centering}
%     \caption{Classification results II} 
%     \label{table:sed_results}
%     \begin{tabular}{ | c | l | l |}
%         \hline
%         Classifiers & ROC AUC & std \\ \hline
%         2D-CNN (k = 3) + 2SAP & 0.8009& 0.1009\\
%         2D-CNN (k = 3) + 2SAP & 0.8248& 0.0790\\
%         2D-CNN (k = 3) + 1SAP + 2D-CNN (k = 3) + 1SAP  & 0.8330& 0.0745\\
%         2D-CNN (k = 5)+ 1SAP + 2D-CNN (k = 3) + 1SAP  & 0.8520& 0.0577\\
%     \hline
%     \end{tabular}
% \end{table*}

% \begin{table*}[!t]
% \small
%     \centering
%     % \captionsetup{justification=centering}
%     \caption{Vowel specific } 
%     \label{table:sed_results}
%     \begin{tabular}{| c | l l |}
%         \hline
%         CNN & ROC AUC & std \\ \hline
%         Vowel a &0.5703	&0.1188 \\
%         Vowel i & 0.8391 &0.1022\\
%         Vowel u &0.8962	&0.0674\\
%         Vowel u + a &0.8039	&0.0741\\
%         Vowel a + i &0.8956	&0.0869\\
%         Vowel i + u &0.6905	&0.0642\\
%     \hline
%     \end{tabular}
% \end{table*}

\section{Experiments}
\label{sec:experiments}

\textbf{Dataset: }
We used recordings of extended vowels /a/, /i/, /u/ of COVID-19 positive and negative (medically tested) individuals. The data were collected under clinical supervision and curated by Merlin Inc., a private firm in Chile. Of 512 data instances, we chose 19 recordings that were collected within 7 days of testing. These comprised 10 females (5 positive) and 9 (4 positive) males. The recordings comprised 8kHz sampled signals recorded on commodity devices.

\begin{figure}[!htb]
    \centering
    \includegraphics[width=3.5in]{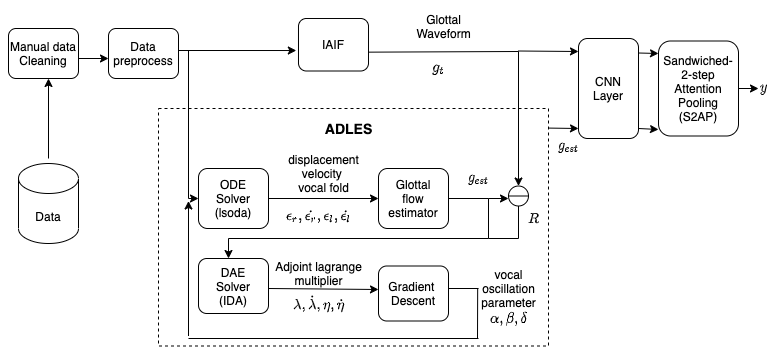}
    \caption{Covid-19 analysis system setup}
    \label{fig:system_Setup}
\end{figure}

\textbf{Setup: }
\label{sec:setup}
Fig. \ref{fig:system_Setup} describes the overall setup used for detection of COVID-19. For each file we use a window $\tau = 50ms$ with shift $o=25ms$, resulting in 3835 frames in all. The extracted $GFW_{filter}$ and $GFW_{model}$ were analyzed with a CNN classifier for the a binary COVID-19 detection task. 3-fold cross validation experiments were performed, where the folds were stratified by speaker: no speaker in the training set was included in the test set, to ensure that the classifier did not learn speaker dependent characteristics. The performance metrics used were the area under curve (ROC-AUC) of the the receiver operating characteristics graph, and its standard deviation (STD). $GFW_{filter}$ was obtained using the IAIF algorithm \cite{alku1992glottal} and $GFW_{model}$ was obtained using the ADLES algorithm.

\vspace{-0.1in}
\section{Results}
\label{sec:results}

\textbf{Classifier results: }
Table \ref{table:class_performance} shows the classification results in terms of ROC-AUC (averaged) and STD in the 3-fold cross-validation experiment. The classifier used was a convolutional neural network (CNN).

\begin{table}[!htb]
      \centering
        \begin{tabular}{ | c | l | l | l |}
        \hline
        Feature extractor & Pooling & ROC-AUC & STD \\ \hline
        - &2AP & 0.6611 & 0.0978\\
        - &2SAP & 0.7925 & 0.1073\\
        CNN (1,3,32) & 2AP & 0.8009& 0.1009\\
        CNN (2,3,32) & 2AP & 0.8248& 0.0790\\
        CNN (2,3,32) & 2SAP  & 0.8330& 0.0745\\
        CNN (2,5,64) & 2SAP  & \textbf{0.8520}& \textbf{0.0577}\\
    \hline
    \end{tabular}
    \caption{Classifier performance on 3-fold cross validation}
    \label{table:class_performance}
		\vspace{-0.1in}
\end{table}

\begin{table}[!htb]
\centering
\begin{tabular}{ | c | c | c | c|c|c|c|}
\hline
 & /a/ & /i/ & /u/ & /a/+/i/ & /a/+/u/ & /i/+/u/ \\
\hline
AUC & 0.57 & 0.839 & 0.896 & 0.690 & 0.804 & \textbf{0.900} \\
STD & 0.119 & 0.102 & 0.067 & 0.064 & 0.074 & \textbf{0.062} \\
\hline
\end{tabular}
\caption{Performance of best model on individual extended vowels and their combinations}
\label{table:each_vowel}
\end{table}

We initially use no feature extractors, and directly do a 2-step attention pooling (2AP). This achieves an ROC-AUC of 0.66, indicating the presence of large differences (lack of one-to-one correspondence) between $GFW_{filter}$ and $GFW_{model}$. This informs us that  neighboring time instances must also be considered. We therefore introduce a new CNN layer with a kernel size of 3 to capture the immediate neighborhoods samples in the frame. In the table \ref{table:class_performance}, the label  CNN (1,3,32) indicates 1 CNN layer with kernel size 3 and filter size 32. This improves the ROC-AUC by 21\% over the best performance without this layer. 

Visualizing the attention patterns reveals that this layer highlights the phase difference and residual between $GFW_{filter}$ and $GFW_{model}$. To better detect the patterns temporally (along the time-axis), we then introduce a second CNN layer to increase the receptive field (denoted by 2,3,32/2AP in Table \ref{table:class_performance}). This is known to help capture \textit{long term} patterns. This results in a further 1\% relative improvement.  

As explained in Sec. \ref{sec:attention}, we now introduce 2SAP (denoted by 2,3,32/2SAP in Table \ref{table:class_performance}). This improves performance by ~19.8\% compared to the standard 2AP, achieving an ROC-AUC of 0.833 when coupled with the initial feature extractor. The final model (kernel size 5, no. of filters 64, denoted by 2,5,64/2SAP) further increases the receptive field \cite{Purwins_2019} and achieves our benchmark ROC-AUC of 0.852. %The ROC-AUC curves of the best model across each fold are shown in figure \ref{fig:CNN_all_folds_auc_roc}.

Note that the model used here is relatively simple compared to the deep stacked CNNs used in other speech and audio classification tasks. Deep-stacking the CNN layers may improve the performance further, but CNN architectures are not focus of  this paper. By using a couple of simple CNN layers here, we show the utility of the proposed methodology in identifying the features that best capture the anomalies in vocal fold motion in a manner that optimizes classifier performance for the current task (detection of  COVID-19). 

\textbf{Vowel analysis: }
Table  \ref{table:each_vowel} shows the results obtained when our best model  derived above (CNN(2,3,64)/2SAP) is trained on particular vowels. The best performance achieved (ROC-AUC of 0.9) is  for the combination of vowels /u/ (high back vowel) and /i/ (high front vowel). The vowel /a/ (low back vowel) exhibits the least importance in this task. This seems to indicate that the vocal folds are unable to vibrate such that higher harmonics have normal energy distributions. However, this conjecture remains to be validated using actual measurements of vocal fold movements.

% \begin{figure}
%     \centering
%     \includegraphics[width=2.5in, keepaspectratio]{figs/cnn_all_folds.png}
%     \caption{Auc roc curve for CNN + 2SAP model for each fold}
%     \label{fig:CNN_all_folds_auc_roc}
% \end{figure}

\begin{figure}
    \centering
    \includegraphics[width=3.3in]{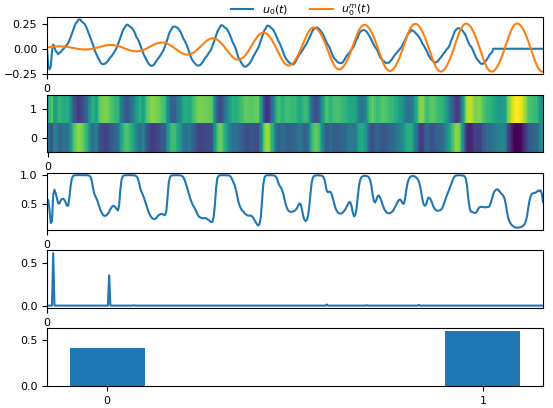}
    \caption{Visualization of attention weights and in the Sandwiched 2-step attention pooling scheme}
    \label{fig:attention_plot}
\vspace{-0.1in}
\end{figure}

\textbf{Interpretability: }
A key advantage of using S2AP is its ability to visualize the contribution of each feature and each time step to classifier's task (the binary task of detecting the presence of COVID-19 in this case). We illustrate this with the example in Fig. \ref{fig:attention_plot}, that shows the attention analysis of a randomly selected frame (a 50ms window) from a COVID-19 positive patient. Going from top to bottom in this figure, panel 1 shows the feature streams: $GFW_{filter}$ and $GFW_{model}$ for this frame. Initially, $GFW_{model}$ is not able to match $GFW_{filter}$. 
Panel 2 shows the attention weights from the $1^{st}$ step of the proposed method. The features are on $y$-axis and time is on the $x$-axis. The feature stream labeled corresponds to  $GFW_{model}$ and the one labeled 0 corresponds to $GFW_{filter}$ estimated using IAIF. As we see, the $1^{st}$ step attention alternately focuses on feature 0 and 1 to detect whether the two features match or not, i.e. on the residual as a function of time. In this panel the lighter shades indicate higher values of the attention weights.

The output of $1^{st}$ step attention is visualized in the third panel. It is high when the peaks and lows of two waves are in synchrony, and can be thought of as detecting phase-locking of the two waves, weighted by the attention. The fourth panel shows the time-level ($2^{nd}$ step) attention weights. The $2^{nd}$ step attention output peaks at the initial point where two GFWs not synchronized. This is the time that contributes the most  significant information for the classification task. The fifth panel is a histogram with 2 bins, centered at 0 and 1. This corresponds to the  output of the binary classifier obtained after $2^{nd}$ step attention. The classifier gives higher weight to ($>$ 0.5) to class 1, which indicates the presence of COVID-19 correctly in the chosen frame, as confirmed by its true label.

Being able to visualize what the classifier focuses in this task  paves the way for further improving it by examining its mistakes. It also provides human interpretability -- a way of adding expert human intervention to black-box model decisions. For example, if the model uses an erroneous pattern leading to decision of COVID-19 traces in the data, it can be discarded as a feature if the human expert does not agree with what entity the attention is focusing on.

\vspace{-0.1in}
\section{Conclusions}
\label{sec:conclusion}
For the detection of COVD-19 from voice, we introduce a method for identifying meaningful features from a comparison of the GFW from inverse filtering to that obtained by solving a dynamical system model for vocal fold oscillations. Our method uses a novel attention mechanism (Sandwiched 2-step Attention Pooling (S2AP)) built within a CNN framework to identify the most discriminative features. For this task it shows that the residual and the phase difference between the two GFWs are the most promising features. Using these for COVID-19 detection yields an ROC-AUC of 0.9 when used with the extended vowels /u/ + /i/.
Overall, this methodology paves the way for comparing any set of feature streams for feature discovery, and can potentially be used in any general classification task where the comparison of two feature streams is expected to yield meaningful features.

\vfill\pagebreak

\bibliographystyle{IEEEbib}
\bibliography{refs}

\end{document}